# Arithmetic with optical topological charges in step-wise excited Rb vapour

Alexander M. Akulshin,[1,*] Irina Novikova,[2] Eugeniy E. Mikhailov,[2] Sergey A. Suslov,[3] and Russell J. McLean[1]

[1] Centre for Quantum and Optical Science, Swinburne University of Technology, Melbourne, Australia
[2] College of William & Mary, Williamsburg, Virginia 23185, USA
[3] Department of Mathematics, Faculty of Science, Engineering and Technology, Swinburne University of Technology, Melbourne, Australia

*Corresponding author: aakoulchine@gmail.com

We report on experimentally observed addition, subtraction and cancellation of orbital angular momentum (OAM) in the process of parametric four-wave mixing that results in frequency up- and down-converted emission in Rb vapour. Specific features of OAM transfer from resonant laser fields with different optical topological charges to the spatially and temporally coherent blue light have been considered. We have observed the conservation of OAM in nonlinear wave mixing in a wide range of experimental conditions including a non-collinear geometry of the applied laser beams and furthermore that the coherent blue light accumulates the total orbital angular momentum of the applied laser light. Spectral and power dependences of vortex and plane wave front blue light beams have been compared.

Structured light [1,2] is a rapidly emerging subfield of optics. Optical vortices are of great interest owing to their numerous prospective applications including optical free space and fibre communication [3,4], particle manipulation [5], and quantum information processing with novel quantum information protocols [6].

Vortex, or twisted, light is characterized by a helical phase front. A vortex beam usually has an annular ring intensity profile with a phase singularity at the beam centre that results in vanishing on-axis light intensity. Analytically, a vortex beam could be characterized based on paraxial eigensolutions of the wave equation in cylindrical coordinates, i.e. the Laguerre-Gaussian (LG) modes. This doughnut-type intensity cross section makes twisted laser beams ideal for optical trapping and optical tweezers [7].

Every photon of vortex light carries $|\ell|\hbar$ orbital angular momentum (OAM), where integer $\ell$ called the optical topological charge is equal to the number of twists in a wave front per unit wavelength. This angular momentum could be large with $|\ell|>>1$ in contrast to the spin angular momentum value of $\pm\hbar$ for circularly polarized light. Positive and negative values of $\ell$ correspond to clockwise and counter clockwise rotation of the wave vector around the beam axis, respectively, while $\ell=0$ describes a beam with plane phase front. OAM of light could transfer torque to a microscopic object as was demonstrated in [8].

Much attention has been also given to the study of soliton-type vortex light propagation in nonlinear bulk and atomic media [9]. One of the first demonstrations of OAM transfer from the applied laser light to a generated optical field of similar wavelength was realized via parametric wave mixing in cold Cs atoms [10]. Later, it was proven that the optical topological charge $\ell$ is conserved during a nonlinear process. The conservation and entanglement of OAM in a parametric frequency down-conversion process was demonstrated in [6], and the storage of OAM of light and its subsequent retrieval has been achieved in warm atomic vapours [11] and in an ensemble of cold atoms [12]. Recently, conservation of OAM in degenerate four-, six- and eight-wave mixing was reported in a laser cooled atomic sample [13]. The transfer of OAM from laser light to optical fields from different spectral regions generated in atomic media due to parametric four wave mixing (FWM) was also reported [14,15]. Optical topological charge conservation was used to distinguish nonlinear processes in atomic media [15].

Here we report experimental observations of optical topological charge transfer from the applied laser radiation to almost frequency-doubled, spatially and temporally coherent blue light (CBL) generated in Rb vapour as a result of a parametric FWM process [16,17,18,19]. The optical topological charge of the frequency up-converted light is related to the applied laser light charges via simple arithmetical operations, such as addition, subtraction and multiplication. Better understanding of the OAM transfer from laser light to frequency converted radiation is particularly important for preparing entangled light fields from different spectral regions.

The coherent blue and mid-IR light generation occurs in atomic Rb vapour stepwise excited to the $5D_{5/2}$ level as illustrated in Fig. 1a. A schematic of our experimental setup for studying OAM transfer from the applied laser light to the frequency up-converted radiation depicted in Fig. 1b is similar to that previously described in [15,19].

The resonant radiation is provided by two extended cavity diode lasers tuned to the $^{85}$Rb $5S_{1/2}(F=3)$-$5P_{3/2}$ and $5P_{3/2}$-$5D_{5/2}$ transitions at 780 and 776 nm respectively, that are linearly polarized in the same direction, and combined on a non-polarizing beam-splitter. Typically, the angle between the two laser beams does not exceed 0.5 mrad, although in some experiments the laser beams are crossed inside the cell at larger angles, up to 14 mrad.

The combined bi-chromatic laser beam is sent through a polariser, a quarter-wave plate and a Rb vapour cell. In this work we use a heated 5-cm long glass cylindrical cell containing Rb vapour with natural isotopic abundance and without buffer gas. The maximum laser powers at 780 and 776 nm before entering the cell are 10 and 5 mW, respectively. In most cases the applied laser light is gently focussed to a waist of ~100 µm inside the Rb cell. The cell axis is intentionally tilted with respect to the direction of the combined laser beam by approximately 100 mrad to minimize a cell orientation effect that will be discussed elsewhere.

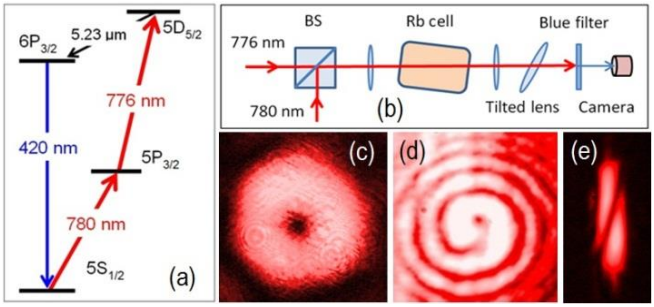

Fig. 1. (a) Schematic of Rb atom energy levels involved in step-wise excitation and subsequent FWM process. (b) Sketch of the experimental setup. (c) Intensity cross section of single-charged vortex laser light ($\ell$ = +1) taken by CCD. (d) Spiral-type interference pattern of single-charged vortex and plane wavefront laser beam at 776 nm. (e) Tilted lens profile of single-charged vortex laser light.

A specially designed resistive heater for the cell provides the required temperature gradient to prevent Rb atoms from condensing on the cell windows. The Rb atom number density evaluated by using experimentally recorded absorption profiles on the Rb D2 line and specialized software [20], is kept within the range $(0.05 - 2) \times 10^{12}$ cm$^{-3}$. At such number densities collisional broadening [21] is negligible compared to the natural linewidth of optical transitions involved in the step-wise excitation and cascade decay.

The optical frequency of the first-step laser at 780 nm is either locked to a Doppler-free dispersion-shaped polarization resonance obtained in an auxiliary Rb cell on the $^{85}$Rb $5S_{1/2}(F=3)$-$5P_{3/2}(F'=4)$ transition or scanned across the Doppler broadened absorption profile on the $5S_{1/2}(F=3)$-$5P_{3/2}$ transition. The frequency of the second-step laser at 776 nm is tuned to maximize the CBL power.

Vortex laser light is created from plane wave front laser light using a phase mask or a blazed diffraction grating with a fork-type singularity. The phase mask for single-charged vortex generation is an optical spiral staircase structure with eight radial sectors that progressively build the total phase difference of $2\pi$ after a complete rotation of the azimuthal angle. The transparency of the single-charged vortex light phase mask is about 95%.

Figure 1c shows the intensity profile of laser light transmitted through the phase mask as recorded by a camera. The zero intensity at the beam core suggests a phase singularity. However a doughnut-type intensity profile is a necessary but not sufficient condition for an optical beam to be vortex-bearing. Observing a spiral-shaped interference pattern produced by mixing with a plane wavefront reference beam (Fig. 1d) is the most convincing way to demonstrate non-zero topological charge of an optical beam. However, in the case of CBL this technique is not straightforward as reference blue light produced independently even by the same lasers might have slightly different optical frequencies [22] preventing the observation of a stationary interference pattern. The self-interference method, in which a fraction of the probed light is expanded and mixed with an unexpanded fraction to produce interference patterns, can also produce confusing results as the appearance and visibility of characteristic spiral- or fork-type structures are very sensitive to the alignment. Other methods exist for measuring the optical topological charge or OAM quantum number [1, 2], but we find that in our case the tilted convex lens method [23], which relies on astigmatic transformation of an optical vortex beam in the vicinity of the focal plane [24], is the most suitable approach because of its simplicity and robustness. The number of high-contrast dark stripes across vortex beam image produced by a tilted lens is equal to the topological charge carrying by the beam, as Figure 1e illustrates.

First we consider the situation when FWM occurs with one twisted and one plane phase front laser beam. Typical images of CBL produced by a single-twisted 776 nm and plane phase front 780 nm laser beam are shown in Fig. 2. In the case of perfectly overlapping laser beams the experimentally observed CBL doughnut-shaped intensity cross section (Fig. 2a) looks similar to the theoretical $LG_0^1$ Laguerre-Gaussian mode. The topological charge of the CBL is revealed by its transformed image obtained with a tilted lens (Fig. 2b). This image is very similar to the result shown in Figure 2c of our numerical modelling, which follows the methodology of Ref. [23], of the intensity profile of a single-charged vortex beam in the vicinity of the focal plane of the tilted lens. A single high-contrast dark stripe appearing across the images indicates that the CBL beam carries a single topological charge. This means that the single unit of OAM has been transferred from the laser light at 776 nm to the blue light in the FWM process. If the first-step laser light at 780 nm carries a single topological charge instead of the 776 nm laser light, OAM also transfers to the CBL.

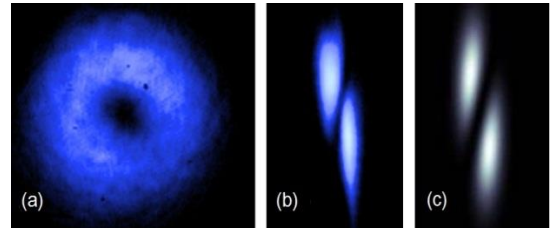

Fig. 2. Intensity profiles of coherent blue light produced with single-charged vortex light at 776 nm ($\ell$=+1) and plane wavefront light at 780 nm ($\ell$=0) without (a) and (b) with lens tilting, respectively. FWM occurs cell containing saturated Rb vapour with approximately $N \approx 0.5 \times 10^{12}$ cm$^{-3}$ Rb atom number density. (c) Calculated tilted lens profile of single-charged vortex light ($\ell$ = +1).

We have repeated this experiment over the wide range $(0.05-2) \times 10^{12}$ cm$^{-3}$ of Rb number densities as well as under more than 50-fold laser intensity variations. We find that the transverse intensity profile of the generated CBL is a complex function of alignment, laser frequency detuning, polarizations and intensities of the applied laser fields as well as the atom number density $N$ and the magnetic field environment. Nevertheless, the high-contrast dark stripes across the tilted lens CBL images are always present confirming that the blue light accumulates the total topological charge of the laser light.

The transfer of OAM from laser light to CBL occurs even if the applied laser beams are not perfectly collinear. In this case the direction of vortex CBL determined by the phase matching condition ($\boldsymbol{k_1} + \boldsymbol{k_2} = \boldsymbol{k_{IR}} + \boldsymbol{k_{BL}}$, where $\boldsymbol{k_1}$, $\boldsymbol{k_2}$, $\boldsymbol{k_{IR}}$ and $\boldsymbol{k_{BL}}$ are the wave vectors of the light beams at 780 nm, 776 nm, 5.23 µm and 420 nm, respectively) does not coincide with the direction of either of the laser beams.

Figure 3a shows intensity profiles of the applied laser beams at 780 and 776 nm that intersect inside the Rb cell at 6 mrad. The CBL images in Figure 3 (b, c) are remarkably similar to the images obtained for the collinear geometry discussed above. The high-contrast dark stripe convincingly demonstrates that the $\ell$ = +1 topological charge has been transferred to CBL despite the angle between the laser beams.

OAM transfer occurs even at larger crossing angles. Figure 3d shows well-separated tilted lens images of both laser beams and CBL detected simultaneously. The characteristic high-contrast tilted dark stripes confirm the transfer of a single topological charge to the CBL.

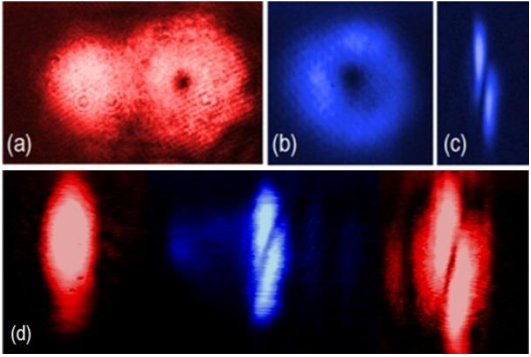

Fig. 3. a) Intensity profiles of the applied single-charged vortex laser light at 776 nm and plane wavefront light at 780 nm crossed inside the Rb cell with atom number density N ≈ 3.5 × 10$^{11}$ cm$^{-3}$. The laser beam intersection angle inside the cell is 6 mrad. (b, c) Intensity profiles of blue light generated by non-collinear laser beams without and with lens tilting, respectively. (d) Vortex blue light produced by vortex light at 776 nm and plane wavefront light at 780 nm crossed inside the Rb cell at 14 mrad that is near the limit imposed by the design of the cell heater.

We note that a violation of the total OAM conservation in the case FWM of non-collinear vortex beams in a bulk medium was recently reported [25]. Although such violation to our knowledge has not been observed in atomic media, the peculiarities of OAM transfer from the non-collinear vortex laser light to the product of parametric FWM require further closer examination. Of particular interest is the transition to conical emission that is common for parametric FWM [26, 27], as well as an evaluation of the critical angle for the OAM conservation and other restrictions imposed by the applied laser light geometry.

In the previous experiment on OAM transfer to blue light [14] a single spatial light modulator was used for applying the same topological charge simultaneously to both laser fields. In our study, independently prepared vortex light at 776 and 780 nm provides new opportunities for analysing the details of the optical topological charge transfer. Results of the following experiments, for example, could be considered as addition and subtraction of OAM in the process of parametric FWM.

We first produce a single-charged vortex beam at 780 nm using the forked grating ($\ell_{780}$=+1), combine this beam with the plane phase front light at 776 nm ($\ell_{776}$ =0), then send the combined bi-chromatic output through the single-charge phase mask, and finally compare the topological charges of CBL generated in the case of two opposite orientations of the mask.

In one orientation, the phase mask adds one unit of a topological charge ($\ell$=+1) to both the transmitted laser fields, making the 776 and 780 nm laser light single- and double-charged, respectively. Thus, the applied laser light carries three units of OAM, since $\ell = \ell_{780} + \ell_{776}$ =3. Three clearly resolved dark strips across the tilted lens image (Fig. 4a) demonstrate that this total OAM has been accumulated by blue light.

The oppositely oriented phase mask in the combined beam adds one unit of negative topological charge ($\ell$ =-1) to both laser fields, producing OAM compensation for the 780 nm laser light ($\ell_{780}$ =0) and single-charged vortex at 776 nm ($\ell_{776}$=-1). Such laser light generates vortex CBL with the opposite chirality or negative topological charge ($\ell_{CBL}$=-1) as evidenced by the single dark stripe with the opposite inclination across the image show in Fig. 4b. No phase mask in the combined bi-chromatic beam results in CBL with single positive topological charge ($\ell_{CBL}$=$\ell_{780}$=+1) as Figure 4c demonstrates. The observed accumulation on total OAM of the applied laser radiation by the frequency up-converted light was attributed in [14] to the spatial overlap with the laser fields being better for the CBL than for 5.23 μm light.

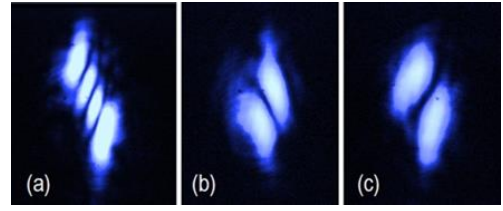

Fig. 4 (a, b, c). Images of CBL with different total topological charges ($\ell_{CBL}$ =+3, $\ell_{CBL}$ =-1 and $\ell_{CBL}$ =+1, respectively) observed using the tilted lens method.

We also find that the applied vortex laser light can produce OAM-free CBL if the equally-charged vortices at 780 and 776 nm have opposite handedness independently produced by the forked grating and the phase mask before combining on the beam splitter. The observed cross-section of the intensity of blue light generated by the single-charged vortex laser light at 780 and 776 nm with opposite chirality is shown in Fig. 5a. Here the doughnut-type intensity profile of the CBL results from the zero on-axis intensity in both applied laser beams. The tilted lens image in Fig. 5b does not contain a characteristic dark stripe. Instead, it closely resembles the calculated image of a vortex-free beam with a doughnut-type intensity distribution (Fig. 5c). That confirms that such pseudo-vortex light does not carry any topological charge despite its doughnut-type intensity profile. Figure 5d demonstrates the tilted lens image of CBL generated under exactly the same experimental conditions but with the same handedness of single-charged vortex laser light at 776 and 780 nm ($\ell_{780}$= $\ell_{776}$ =+1). Two high-contrast dark stripes across the tilted lens image indicate that in this case the frequency up-converted light carries two units of OAM, accumulating the total topological charge of the applied laser light. The calculated tilted lens profile of a double-charged vortex beam (Fig. 5e) is consistent with the experimentally observed image.

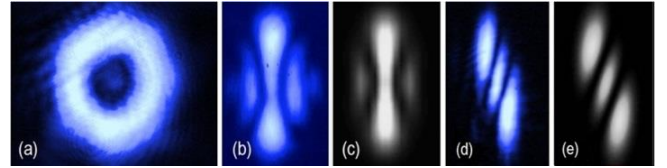

Fig. 5. (a) Intensity profile and (b) tilted lens image of CBL generated with laser fields having single topological charge of opposite helicity ($\ell_{780}$ =+1 and $\ell_{776}$= -1). (c) Calculated tilted lens image of vortex-free beam with doughnut-shaped intensity cross section (pseudo-vortex light). (d) Tilted lens profile of CBL generated with the same handedness of single-charged vortex laser light at 776 and 780 nm ($\ell_{780}$= $\ell_{776}$ =+1). (e) Calculated tilted lens profile of double-charged vortex light.

Finally, we compare power and spectral characteristics of the vortex and vortex-free CBL generated in Rb vapours. To eliminate some effects attributed to different spatial characteristics of the applied laser beams at 780 and 776 nm, e.g. different divergence, astigmatism, diameter and their mutual overlapping, the combined beam is first transmitted through a single-mode fibre making spatial properties of the laser fields at 780 and 776 nm identical.

Figure 6a shows typical dependences on the 780 nm laser frequency detuning of plane phase front and vortex CBL generated in the Rb cell under the same experimental conditions. The spectral dependences are significantly different in amplitude and width. It appears that the vortex CBL is generated over a significantly narrower range of 780 nm laser frequencies than for plane phase front light. We also find that vortex CBL has a higher threshold and its power grows more slowly compared to the plane wavefront blue light (Fig. 6b).

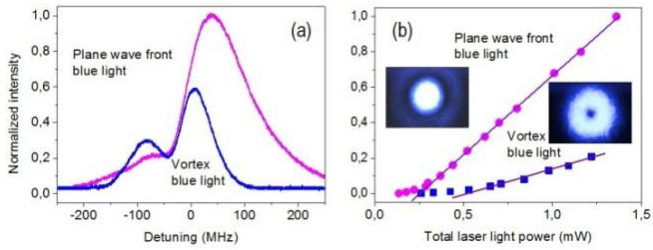

Fig. 6. a) Intensity of plane wave front and double-charged vortex blue light as a function of frequency detuning of the 780 nm laser from the $^{85}$Rb $5S_{1/2}$(F=3)-$5P_{3/2}$(F'=4) transition. b) Power dependence of plane wave front and double-charged blue light.

We attribute these effects of intensity reduction and spectral profile narrowing to less efficient excitation of the Rb atoms. Indeed, the peak intensity of the vortex laser light is significantly lower than that of the plane wavefront light because of increased cross section and approximately 5% intensity reduction due to transmission through the phase mask. This could be also due to the fact that the pencil-shaped excitation region formed by vortex laser light inside the Rb cell contains no excited atoms along the longitudinal axis making the process of ASE less efficient, leading to a higher threshold for 5.23 μm field, and its lower intensity in turn increases the threshold and reduces the intensity of the CBL.

We also note that the optimal frequency tuning of the applied laser fields is different for vortex and plane wavefront CBLs. Detailed analysis of spectral and intensity properties of CBL will be the subject of a separate paper.

In summary, the transfer of optical topological charge from resonant laser light to temporally and spatially coherent frequency up-converted emission generated in step-wise excited Rb vapours has been experimentally investigated. The topological charge of the generated collimated blue light has been measured by the tilted lens method. The ability of this method to distinguish between real and pseudo optical vortex beams having similar doughnut-like intensity profiles has been confirmed by our numerical modelling and experimental observations.

It has been found that the coherent blue light produced by parametric FWM accumulates the total orbital angular momentum of the applied laser light for a wide range of experimental parameters, such as 50-fold variations of the applied laser power, Rb number density variations in the range (0.3–1.5) × $10^{12}$ cm$^{-3}$ and non-collinear geometry of the applied laser beams. Our search for experimental conditions under which the optical topological charge carried by the applied laser light could partially or completely transfer to the frequency down-converted radiation instead of the frequency up-converted emission, which is important for entangled light generation from different spectral regions will be extended to more symmetric diamond-type energy configurations in alkali vapours.

Spectral and power dependences of vortex and plane phase front blue light reveal that the vortex laser light generates less intense CBL over a narrower spectral range compared to plane phase front light. We explain this fact by less efficient step-wise excitation of Rb atoms and weaker amplified spontaneous emission in the region along the longitudinal axis with no excited atoms. Our experimental findings of OAM conservation in nonlinear frequency conversion form an important step towards the generation of the topologically correlated optical fields from different spectral regions.

**Acknowledgment**. We thank A. Sidorov for useful discussions and Grover Swartzlander for the loan of the vortex phase mask.

## References

1. M. S. Soskin and M. V. Vasnetsov, Prog. Opt. **42**, 219 (2001).
2. A. Yao, and M. Padgett, Adv. Optics and Photonics **3**, 161 (2011).
3. G. Gibson, J. Courtial, M.J. Padgett, M. Vasnetsov, V. Pasko, S.M. Barnett, and S. Franke-Arnold, Opt. Express **12**, 5448–5456 (2004).
4. N. Bozinovic, Y. Yue, Y. Ren, M. Tur, P. Kristensen, H. Huang, A. E. Willner, and S. Ramachandran, Science **340**, 1545 (2013).
5. M. Padgett and R. Bowman, Nat. Photonics **5**, 343 (2011).
6. A. Mair, A. Vasiri, G. Weihs, and A. Zeilinger, Nature **412**, 313 (2001).
7. J. E. Curtis and D. G. Grier, Phys. Rev. Lett. **90**, 133901 (2003).
8. H. He, M. E. J. Friese, N. R. Heckenberg, and H. Rubinsztein-Dunlop, Phys. Rev. Lett. **75**, 826 (1995).
9. A. Desyatnikov, Y. Kivshar, and L. Torner, Prog. Opt. **47**, 291 (2005).
10. J. W. R. Tabosa and D. V. Petrov, Phys. Rev. Lett. **83**, 4967 (1999).
11. R. Pugatch, M. Shuker, O. Firstenberg, A. Ron, and N. Davidson, **98**, 203601 (2007)
12. D. Moretti, D. Felinto, and J. Tabosa, Phys. Rev. A **79**, 023825 (2009).
13. R. A. de Oliveira, G. C. Borba, W. S.Martins, S. Barreiro, D.Felinto, and J.W. R. Tabosa, Opt. Lett. **40**, 4939 (2015).
14. G. Walker, A. S. Arnold, and S. Franke-Arnold, Phys. Rev. Lett. **108**, 243601 (2012).
15. A. M. Akulshin, R. J. McLean, E. E. Mikhailov, and I. Novikova, Opt. Lett. **40**, 1109 (2015).
16. A. S. Zibrov, M. D. Lukin, L. Hollberg, and M. O. Scully, Phys. Rev. A **65**, 051801 (2002).
17. A. Vernier, S. Franke-Arnold, E. Riis, and A. S. Arnold, Opt. Express **18**, 17020 (2010).
18. J. F. Sell, M. A. Gearba, B. D. DePaola, and R. J. Knize, Opt. Lett. **39**, 528 (2014).
19. A. Akulshin, D. Budker, and R. McLean, Opt. Lett. **39**, 845 (2014).
20. Rochester Scientific; http://rochesterscientific.com/ADM.
21. A. M. Akulshin A. A. Celikov, V. A. Sautenkov, T. A. Vartanian and V. L. Velichansky, Opt. Comm. **85,** 21 (1991).
22. A. M. Akulshin, Ch. Perrella, G.-W. Truong, R. J. McLean, and A. Luiten, J. Phys. B: At. Mol. Opt. Phys. **45,** 245503 (2012).
23. P. Vaity, J. Banerji, R.P. Singh, Phys. Lett. A **377** 1154 (2013).
24. A. Ya. Bekshaev, M.S. Soskin and M.V. Vasnetsov, Opt. Commun. **241**, 237 (2004).
25. T. Roger, J. J. F. Heitz, E. M. Wright, and D. Faccio, Sci. Reports **3**, 3491 (2013).
26. W. Chalupczak, W. Gawlik and J. Zachorowski, Opt. Comm. **99**, 49 (1993).
27. A. I. Lvovsky, S. R. Hartmann and F. Moshary, Phys. Rev. Lett. **82**, 4420 (1999).

## References

1. M. S. Soskin and M. V. Vasnetsov, "Singular optics", Prog. Opt. **42**, 219 (2001).
2. A. Yao, and M. Padgett, "Orbital angular momentum: origins, behavior and applications", Adv. Optics and Photonics **3**, 161 (2011).
3. G. Gibson, J. Courtial, M. J. Padgett, M. Vasnetsov, V. Pasko, S. M. Barnett, and S. Franke-Arnold, "Free-space information transfer using light beams carrying orbital angular momentum," Opt. Express **12**, 5448–5456 (2004).
4. N. Bozinovic, Y. Yue, Y. Ren, M. Tur, P. Kristensen, H. Huang, A. E. Willner, S. Ramachandran, "Terabit-Scale Orbital Angular Momentum Mode Division Multiplexing in Fibers", Science **340**, 1545 (2013).
5. M. Padgett and R. Bowman, "Tweezers with a twist", Nat. Photonics **5**, 343 (2011).
6. A. Mair, A. Vasiri, G. Weihs and A. Zeilinger, "Entanglement of the orbital angular momentum states of photons", Nature **412**, 313 (2001).
7. J. E. Curtis and D. G. Grier, "Structure of optical vortices ", Phys. Rev. Lett. **90**, 133901 (2003).
8. H. He, M. E. J. Friese, N. R. Heckenberg, and H. Rubinsztein-Dunlop, "Direct observation of transfer of angular-momentum to absorptive particles from a laser-beam with a phase singularity", Phys. Rev. Lett. **75**, 826 (1995).
9. A. Desyatnikov, Y. Kivshar, and L. Torner, "Optical vortices and vortex solitons", Prog. Opt. **47**, 291 (2005).
10. J. W. R. Tabosa and D. V. Petrov, "Optical pumping of orbital angular momentum of light in cold Cesium atoms", Phys. Rev. Lett. **83**, 4967 (1999).
11. R. Pugatch, M. Shuker, O. Firstenberg, A. Ron, and N. Davidson, "Topological Stability of Stored Optical Vortices", Phys. Rev. Lett. **98**, 203601 (2007).
12. D. Moretti, D. Felinto, and J. Tabosa, "Collapses and revivals of stored orbital angular momentum of light in a cold-atom ensemble", Phys. Rev. A **79,** 023825 (2009)
13. R. A. de Oliveira, G. C. Borba, W. S.Martins, S. Barreiro, D.Felinto, and J.W. R. Tabosa, "Nonlinear optical memory for manipulation of orbital angular momentum of light", Opt. Lett. **40**, 4939 (2015).
14. G. Walker, A. S. Arnold, and S. Franke-Arnold, "Trans-Spectral Orbital Angular Momentum Transfer via Four-Wave Mixing in Rb Vapor", Phys. Rev. Lett. **108**, 243601 (2012).
15. A. M. Akulshin, R. J. McLean, E. E. Mikhailov, and I. Novikova, "Distinguishing nonlinear processes in atomic media via orbital angular momentum transfer", Opt. Lett. **40**, 1109 (2015).
16. A. S. Zibrov, M. D. Lukin, L. Hollberg, and M. O. Scully, "Efficient frequency up-conversion in resonant coherent media", Phys. Rev. A **65**, 051801 (2002).
17. A. Vernier, S. Franke-Arnold, E. Riis, and A. S. Arnold, "Enhanced frequency up-conversion in Rb vapor", Opt. Express **18**, 17020 (2010).
18. J. F. Sell, M. A. Gearba, B. D. DePaola, and R. J. Knize, "Collimated blue and infrared beams generated by two-photon excitation in Rb vapor", Opt. Lett. **39**, 528 (2014).
19. A. Akulshin, D. Budker, and R. McLean, "Directional infrared emission resulting from cascade population inversion and four-wave mixing in Rb vapor", Opt. Lett. **39**, 845 (2014).
20. Rochester Scientific; http://rochesterscientific.com/ADM
21. A. M. Akulshin A. A. Celikov, V. A. Sautenkov, T. A. Vartanian and V. L. Velichansky, "Intensity and concentration dependence of Doppler-free resonance in selective reflection", Opt. Comm. **85**, 21 (1991).
22. A. M. Akulshin, Ch. Perrella, G.-W. Truong, R. J. McLean, and A. Luiten, "Frequency evaluation of collimated blue light generated by wave mixing in Rb vapour", J. Phys. B: At. Mol. Opt. Phys. **45**, 245503 (2012).
23. P. Vaity, J. Banerji, R.P. Singh, "Measuring the topological charge of an optical vortex by using a tilted convex lens", Phys. Lett. A **377** 1154 (2013).
24. A. Ya. Bekshaev, M.S. Soskin and M.V. Vasnetsov, "Transformation of higher-order optical vortices upon focusing by an astigmatic lens", Opt. Commun. **241**, 237 (2004).
25. T. Roger, J. J. F. Heitz, E. M. Wright, and D. Faccio, "Non-collinear interaction of photons with orbital angular momentum", Sci. Reports **3**, 3491 (2013).
26. W. Chalupczak, W. Gawlik and J. Zachorowski, "Degenerate parametric emission in dense barium vapour", Opt. Comm. **99**, 49 (1993).
27. A. I. Lvovsky, S. R. Hartmann and F. Moshary, "Omnidirectional Superfluorescence", Phys. Rev. Lett. **82**, 4420 (1999).